\documentclass[apjl]{emulateapj}
\def\ergs{ergs~s$^{-1}$}
\def\ergcms{ergs$^{-1}$~cm$^{-2}$~s$^{-1}$}
\def\cts{counts~s$^{-1}$}
\def\xa{X41.4$+$60}
\def\xb{X42.3$+$59}
\def\xmm{{\it XMM}}
\def\xmmn{{\it XMM-Newton}}
\def\cha{{\it Chandra}}
\def\xte{{\it RXTE}}
\def\vla{{\it VLA}}

\begin{document}

\title{Origin of the X-ray Quasi-Periodic Oscillations and Identification of a Transient Ultraluminous X-Ray Source in M82}

\author{Hua Feng\altaffilmark{1} and Philip Kaaret}

\affil{Department of Physics and Astronomy, The University of Iowa, Van
Allen Hall, Iowa City, IA 52242}
\altaffiltext{1}{Email: hua-feng@uiowa.edu}

\shortauthors{Feng and Kaaret}
\shorttitle{ULXs in M82}
\submitted{Submitted to ApJ on May 08, 2007}

\begin{abstract}
The starburst galaxy M82 contains two ultraluminous X-ray sources (ULXs), CXOM82~J095550.2$+$694047 (=X41.4$+$60) and CXOM82~J095551.1$+$694045 (=X42.3$+$59), which are unresolved by XMM-Newton. We revisited the two  XMM-Newton observations of M82 and analyzed the surface brightness profiles using the known Chandra source positions. We show that the quasi-periodic oscillations (QPOs) detected with XMM-Newton originate from X41.4$+$60, the brightest X-ray source in M82. Correcting for the contributions of the unresolved sources, the QPO at a frequency of $55.8\pm1.3$~mHz on 2001 May 06 had a fractional rms amplitude of 32\%, and the QPO at $112.9\pm1.3$~mHz on 2004 April 21 had an amplitude of 21\%. The QPO frequency may possibly be correlated with the source flux, similar to the type C QPOs in XTE~1550$-$564 and GRS~1915$+$105, but at luminosities two orders of magnitude higher. X42.3$+$59, the second brightest source in M82, displayed a strikingly high flux of $1.4\times10^{-11}$~ergs$^{-1}$~cm$^{-2}$~s$^{-1}$ in the 2--10 keV band on 2001 May 6. A seven-year light curve of X42.3$+$59 shows extreme variability over a factor of 1000; the source is not detected in several Chandra observations. This transient behavior suggests accretion from an unstable disk.  If the companion star is massive, as might be expected in the young stellar environment, then the compact object would likely be an IMBH.
\end{abstract}

\keywords{black hole physics -- accretion, accretion disks -- X-rays:
binaries -- X-rays: galaxies -- X-rays: individual (M82 X-1, CXOM82~J095550.2$+$694047=X41.4$+$60, CXOM82~J095551.1$+$694045=X42.3$+$59)}

\section{Introduction}
Ultraluminous X-ray sources (ULXs) are nonnuclear, point-like X-ray sources in external galaxies with luminosities (assuming isotropic emission) above the Eddington limit for a 20$M_\sun$ black hole ($3\times10^{39}$ ergs s$^{-1}$). Their fast variability and high luminosities indicate they may contain intermediate-mass black holes \citep[IMBHs;][]{col99,mak00,kaa01}. However, if the emission is beamed \citep{kin01,kor02}, exceeds the Eddington limit \citep{wat01,beg02}, or both \citep{pou07}, IMBHs are not required.

Using {\it ASCA}, \citet{pta99} and \citet{mat99} found that the bright X-ray emission from the central region of the starburst galaxy M82 was highly variable on time scales of hours to days, suggesting it arises from a compact object. The emission from the core of M82 was resolved with \cha\ into several point sources, among which an extremely bright ULX CXOM82~J095550.2$+$694047 was identified at a luminosity of $10^{40}-10^{41}$~\ergs\ \citep{mat01,kaa01}. Following the convention of naming sources in M82 by their offset from $\alpha=09^{\rm h}51^{\rm m}00^{\rm s}$, $\delta=+69\arcdeg54\arcmin00\arcsec$ (B1950), we refer to this source as \xa\ and use the same convention in referring to other X-ray and radio sources in M82. The source is highly variable, indicating it is not a supernova or remnant, and not coincident with the dynamical center of M82, indicating it is not an active galactic nucleus (AGN).  If the radiation is isotropic and Eddington-limited, the source would be an IMBH candidate with a mass in excess of 500$M_\sun$.  The luminosity of \xa\ is too high to be explained by mechanically beamed emission from a stellar mass black hole \citep{kin05}.  An X-ray flare detected using the {\it Rossi X-ray Timing Explorer} (\xte) in February 2005 showed no strong radio emission in contemporaneous {\it Very Large Array} (\vla) observations which ruled out the possibility that the source is relativistically beamed \citep{kaa06}. 

\xmm\ cannot resolve the multiple point sources detected with \cha\ in the central region of M82, but provides superior collection area allowing sensitive studies of timing properties. \citet{str03} analyzed a 27~ks \xmm\ observation of the source obtained in 2001 and discovered quasi-periodic oscillations (QPOs) at a frequency of 54~mHz in the energy range 2--10 keV. The QPOs were confirmed in a 103~ks \xmm\ observation in 2004, however, the frequency shifted to 114~mHz \citep{dew06,muc06}.  The power spectral density (PSD) from the second observation also showed a low frequency break at 34~mHz. \citet{muc06} found that the QPOs changed frequency from 107~mHz to 120~mHz during the second observation, and discovered a plausible harmonic ratio of 1:2:3 for QPO frequencies with more data from \xte. Due to the limited angular resolution of \xmm\ (and \xte), all of these analyzes treated the emission as arising from a single point source, called M82 X-1.

However, higher angular resolution observations made with \cha\ show that M82 contains two ultraluminous X-ray sources, CXOM82~J095550.2$+$694047 (=\xa) and CXOM82~J095551.1$+$694045 (=\xb), and several dimmer sources which contribute to the emission ascribed to M82 X-1. The second ULX, \xb, has shown pronounced variability, appearing as the second brightest source in M82 on 1999 Oct 28, but being undetected on 2000 Jan 20 \citep{mat01,kaa01}.  A Chandra observation in February 2005 revealed \xb\ in a bright state with strong variability, but the sensitivity was not adequate to confirm or reject the presence of QPOs.  

It is of interest to determine which of the two ULXs is the source of the QPO detected with \xmm.  We revisited the two \xmm\ observations of M82 (\S\ref{sec:data}), corrected their astrometry using a \cha\ observation (\S\ref{sec:ast}), resolved the source count rates of \xa\ and \xb\ from surface brightness fits (\S\ref{sec:fit}), and compared them with timing analysis done with various source regions to determine the QPO origin (\S\ref{sec:timing}). A seven-year light curve of \xb\ using all available \xmm\ and \cha\ archival data is presented in \S\ref{sec:x42}. The results are discussed in \S\ref{sec:dis}.

\section{Observations and Data Analysis}
\label{sec:data}
\xmmn\ observations of M82 made on 2001 May 06 (ObsID 0112290201) and 2004 April 21 (ObsID 0206080101) were used for image and timing analysis. The \cha\ observation with the High Resolution Camera (HRC) made on 1999 Oct 28 (ObsID 1411-1) was used to correct the astrometry of \xmm\ images. All \cha\ and \xmm\ observations were used to produce the light curve of \xb. We used SAS 7.0.0 with up to date calibration files for \xmm\ data reduction and CIAO 3.3.0.1 with CALDB 3.2.2 for \cha\ data reduction.

We corrected the \xmm\ astrometry by matching point sources on the \cha\ and \xmm\ images.  Point sources in \xmm\ images were found using {\tt edetect\_chain} in the 2--10~keV band.  Point sources within 18\arcsec\ of \xa\ were not used because the diffuse emission and the two unresolved bright ULXs could bias the \xmm\ source detection.  Positions for \cha\ sources, shown in Table~\ref{tab:src}, were obtained using {\tt wavdetect}, and then corrected by shifting \xa\ to the position $\rm R.A.=09^h55^m50\fs17$, $\rm Decl.=+69\arcdeg40\arcmin46\farcs7$ \citep{kaa01}.  We note that none of our results depend on the absolute astrometry.  We adopt the position of X41.4+60 from \citep{kaa01} as a reference in order to keep the source names consistent with those previously used in the literature.  We found 8 matched point sources between the MOS1 and HRC image for the first observation and 7 for the second observation.  All these sources were used to align the \xmm\ MOS1 images to the HRC image.

\subsection{Astrometry}
\label{sec:ast}

\begin{deluxetable}{cccc}[th]
\tablecolumns{4}
\tablewidth{\columnwidth}
\tablecaption{\cha\ source positions used in the surface brightness fits
\label{tab:src}}
\tablehead{
\colhead{source \#} & \colhead{R.A.} & \colhead{Decl.} & \colhead{name}\\
\colhead{} & \colhead{(J2000)} & \colhead{(J2000)} & \colhead{} \\
\colhead{(1)} & \colhead{(2)} & \colhead{(3)} & \colhead{(4)}
}
\startdata
1 & 09 55 54.68 & +69 41 01.1 &\\
2 & 09 55 52.31 & +69 40 54.1 &\\
3 & 09 55 51.48 & +69 40 36.0 &\\
5 & 09 55 51.05 & +69 40 45.3 & \xb \\
7 & 09 55 50.17 & +69 40 46.7 & \xa \\
9 & 09 55 46.61 & +69 40 41.1 & \\
\enddata

\tablecomments{The first column indicates the source index in \citet{mat01}. All positions are obtained from {\tt wavdetect} and then shifted by matching source 7 to the position $\rm R.A.=09^h55^m50\fs17$, $\rm Decl.=+69\arcdeg40\arcmin46\farcs7$ \citep{kaa01}.}
\end{deluxetable}

\subsection{Surface Brightness Fits}
\label{sec:fit}

We use only \xmm\ MOS1 images for the surface brightness fits.  MOS2 images are not used because the MOS2 point spread function (PSF) is not axisymmetric within a few arcseconds of the core.  PN images are not suitable because its pixel size of 4\arcsec\ slightly undersamples the core and is close to the angular distance between the two ULXs. M82 X-1 is a bright X-ray source, so the MOS1 images with $3\times10^4$ source photons in 2--10 keV band for the first observation and $8\times10^4$ for the second provide adequate statistics.

MOS1 images for the 2--10 keV band were created in a $250\arcsec\times250\arcsec$ region around \xa\ and sampled on a $1\arcsec\times1\arcsec$ grid with events screened by FLAG equal to \#XMMEA\_EM and PATTERN between 0 and 12. The effective exposure for the MOS1 image is 30.0~ks for the first observation and 101.2~ks for the second. The PSF for MOS1 can be well described by a King function and the King parameters are stored in the latest calibration file (e.g., XRT1\_XPSF\_0007.CCF for MOS1 by now). We used the command {\tt calview} with an ``EXTENDED'' accuracy level to produce the on-axis King model PSF at 2~keV for MOS1. The King model parameters vary little with energy\footnote{see \S3 and Figure~2 in \xmmn\ current calibration file release notes 167.}.  Therefore, a monochromatic PSF at 2~keV provides an adequate description of the photon distribution for a 2--10 keV point source image.

Six of the nine sources listed in \citet{mat01} were used in the surface brightness fits. However, the source names in \citet{mat01} do not retain the sub-arcsecond accuracy that is necessary for the fit.  We thus used the positions obtained from {\tt wavdetect} as mentioned in \S\ref{sec:ast} and listed in Table~\ref{tab:src}.

The 2-D surface brightness fits of the \xmm\ MOS1 images were performed with the CIAO program {\tt Sherpa}. The PSF is loaded with a size of $128\arcsec \times 128\arcsec$ around the core. We use $\delta$-functions to model the point sources and a constant to model the background. Point sources are fixed at positions quoted in Table~\ref{tab:src}, and allowed to shift together. 

For the first \xmm\ observation, only two point sources, \xa\ and \xb, were considered in the fit. The model had five free parameters: the pattern shifts $\Delta X$ and $\Delta Y$, source amplitudes for the two ULXs (\xa\ and \xb), and a background amplitude taken as constant across the whole field. We applied the CHI GEHRELS statistics in {\tt Sherpa} because many pixels near the edges of the field contained few photons. The best-fit source position is 0\farcs8 away from the \cha\ position, which implies that the astrometric correction was successful. The resolved count rate is 0.246 \cts\ for \xa\ and 0.395 \cts\ for \xb. The background level is $1.6\times10^{-6}$~\cts~arcsec$^{-2}$. In a 18\arcsec-radius circle, the model and data result in $\chi^2=948$ with 1009 degrees of freedom (dof), indicating an adequate fit.

\begin{deluxetable}{ccccc}
\tablecolumns{5}
\tablewidth{\columnwidth}
\tablecaption{Resolved 2--10 keV count rates of \xa\ and \xb\ from surface brightness fits with {\it XMM-Newton} MOS1 images
\label{tab:img}}
\tablehead{
\colhead{\#} & \colhead{ObsID} & \colhead{\scriptsize \xa\ rate} & \colhead{\scriptsize \xb\ rate} & \colhead{$\chi^2/$dof}\\
\colhead{} & \colhead{} & \colhead{(\cts)} & \colhead{(\cts)} & \colhead{}\\
\colhead{(1)} & \colhead{(2)} & \colhead{(3)} & \colhead{(4)} & 
\colhead{(5)}
}
\startdata
1 & 0112290201 & 0.246$\pm$0.007 & 0.395$\pm$0.008 & 948/1009 \\
2 & 0206080101 & 0.305$\pm$0.003 & 0.064$\pm$0.003 & 1453/1005 \\
\enddata

\tablecomments{
Col.~(1): Observation number.
Col.~(2): {\it XMM-Newton} ObsID.
Col.~(3): Best-fit source count rate of \xa.
Col.~(4): Best-fit source count rate of \xb.
Col.~(5): $\chi^2$ and degrees of freedom in an 18\arcsec-radius region around \xa.
All errors are at 1$\sigma$ level.  There is a systematic uncertainty on the count rate of \xb\ of 5\% in the first observation and 30\% in the second due to very nearby sources, see the text, which is not included in the purely statistical errors quoted in the Table.}
\end{deluxetable}

\begin{figure}[h!]
\centering
\includegraphics[width=0.8\columnwidth]{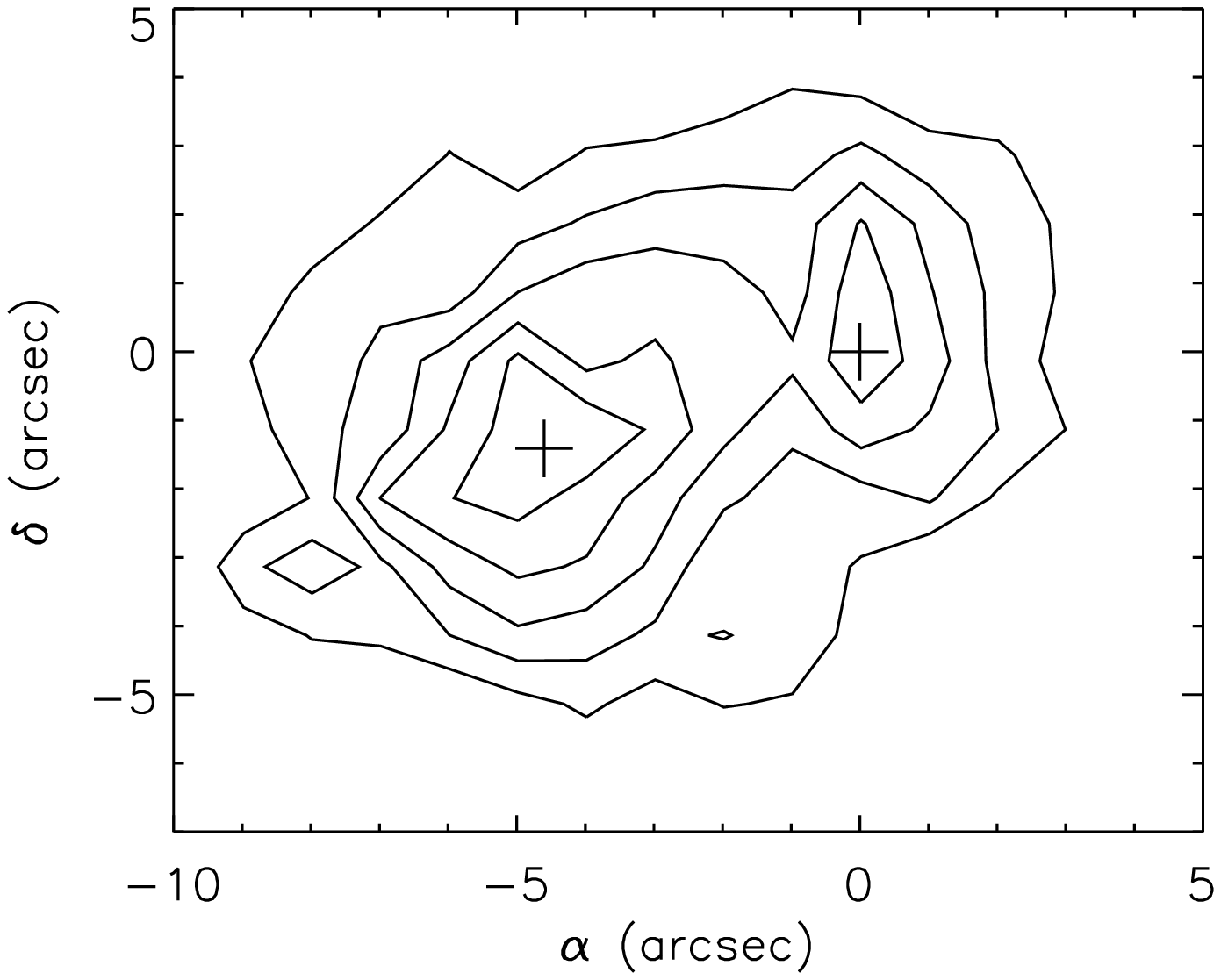} \\
\includegraphics[width=0.8\columnwidth]{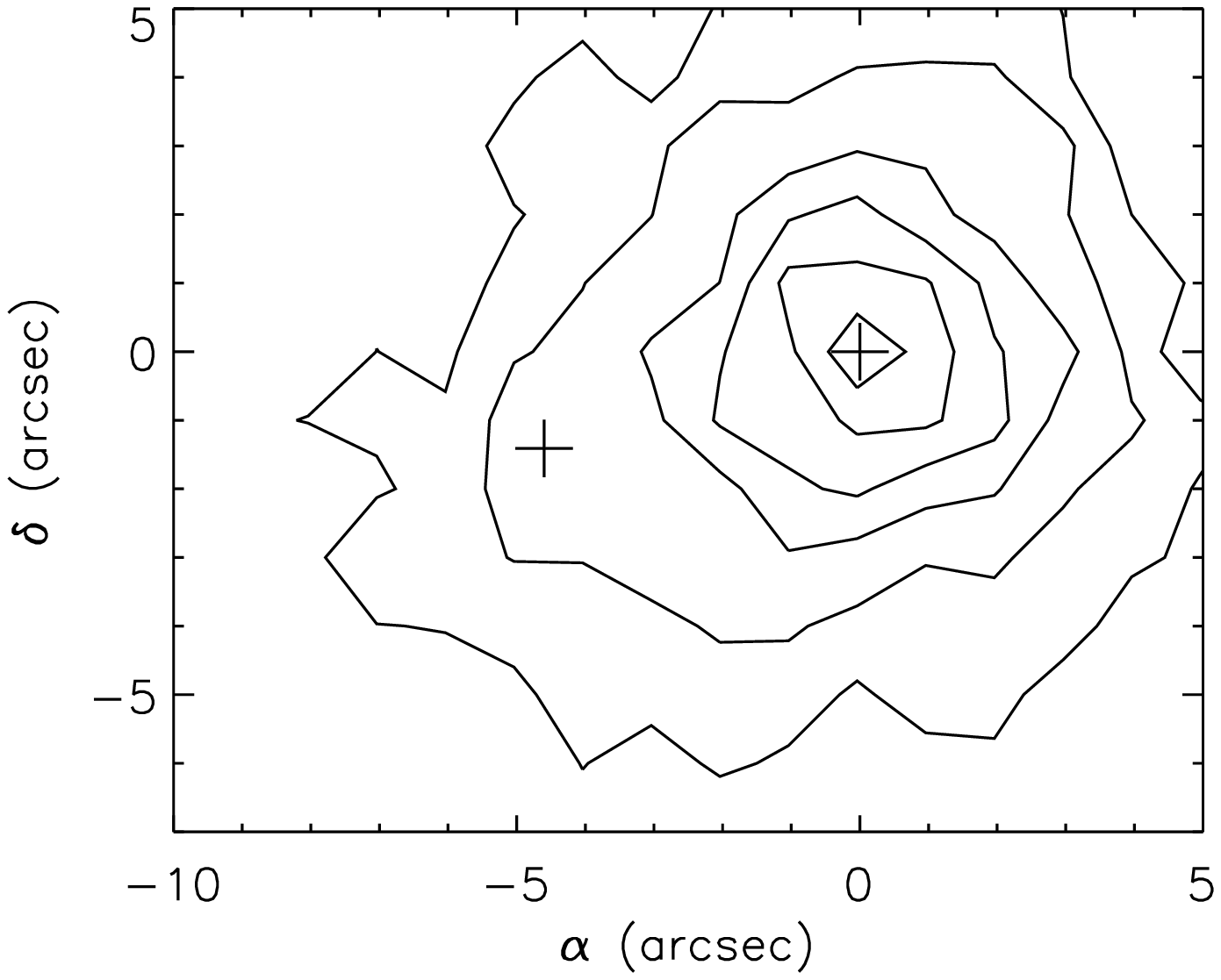} 
\caption{
2--10 keV \xmm\ MOS1 contour maps of the surface brightness in the central M82 region for the first observation ({\it top}) and the second observation ({\it bottom}). Coordinates are relative to the \cha\ position of \xa\ (plus at the origin). The other plus off the origin is the \cha\ position of \xb.  The \xmm\ images have been shifted according to the best-fit positions from the surface brightness profile fitting.
The levels are 1.5, 2.0, 2.5, 3.0 and 3.5 ($10^{-3}$ \cts~arcsec$^{-2}$) for the first observation and 0.7, 1.0, 1.5, 2.0, 2.5 and 3.0 ($10^{-3}$ \cts~arcsec$^{-2}$) for the second.
\label{fig:img}}
\end{figure}

The situation for the second observation is slightly more complex, because the long exposure causes the diffuse emission and other dim sources in the central region to become significant. If we only model \xa\ and \xb\ plus a constant background, the $\chi^2$/dof in the central 18\arcsec\ region is 2230/1009. We thus added another 4 point sources to the model, i.e., sources 1, 2, 3 and 9 in \citet{mat01}. Sources 4, 6 and 8 are not added because source 8 was not bright in the second observation and sources 4 and 6 are too close to \xb\ (source 5).  We then performed the fit with a model including 6 point sources and a constant background with 9 free parameters. The 6 point sources were still set to move together as a whole pattern and their best-fit position is only 0\farcs3 away from the \cha\ position. In this observation, the count rate of \xa\ increased to 0.305~\cts\ while the \xb\ rate decreased to 0.064~\cts. The background level is $3.3\times10^{-6}$~\cts~arcsec$^{-2}$. Modeling the extra 4 sources improved the fit significantly, resulting in $\chi^2/{\rm dof}=1453/1005$. We checked the residual map and found most residuals came from structure in the diffuse emission region which is not included in our model. Even though the fit is not formally adequate, we believe the resolved the count rates are reliable because the emission from the ULXs is much stronger than the diffuse emission.  

After checking all \cha\ ACIS (Advanced CCD Imaging Spectrometer) observations, we found that the count rate in the 2--8 keV band of source 4 varied by a factor of 2, while source 6 appeared constant.  We estimate that sources 4 and 6 contribute a systematic error to the count rate estimated for \xb\ of 5\% in the first observation and 30\% in the second.

Contour maps of the surface brightness (corrected with exposure map and mask) for the two observations are shown in Figure~\ref{fig:img}. From the contour maps we can clearly see that the count rate of \xa\ is slightly lower than \xb\ in the first observation and much higher than \xb\ in the second.  This is consistent with the rates from the surface brightness model fits.

\subsection{Flux from \xa}
\label{sec:flux}

Because \xmm\ is unable to resolve \xa\ from other point sources and the diffuse emission in M82, one needs great caution to make energy spectral analysis with \xmm\ data. So far the only energy spectrum of \xa\ which is neither contaminated by \xb\ nor strongly affected by pileup is from \citet{kaa06} with an off-axis \cha/ACIS observation, in which the spectrum is not contaminated by other sources nor suffers serious and uncorrectable pile-up as found in the on-axis \cha/ACIS observations. In that observation, \xa\ presented a featureless power-law spectrum with a photon index of 1.67. 

We adopt a power-law spectral index of 1.7 \citep{kaa06} with an absorption column density of $3\times10^{22}$~cm$^{-2}$, and use PIMMS to estimate the unabsorbed source flux in the 2--10 keV range according to the resolved count rates (with a factor of 0.7, because PIMMS accepts the counting rate in a 15\arcsec\ region, which contains around 70\% of a 2--10~keV PSF), which gives $8.5\times10^{-12}$~\ergcms\ for the first observation and $1.0\times10^{-11}$~ergs$^{-1}$~cm$^{-2}$~s$^{-1}$ for the second.  The corresponding luminosity is $1.3\times10^{40}$~\ergs\ and $1.7\times10^{40}$~\ergs\ at a distance of 3.63~Mpc \citep{fre94}, respectively, for the first and second observation.  The flux is not very sensitive to the photon index, varying only by 10\% as the photon index varies between 1.5--2.3, but a decrease of the column density to $1\times10^{22}$~cm$^{-2}$ will bring down the unabsorbed flux by 25\%.

\begin{figure}
\centering
\includegraphics[width=0.7\columnwidth]{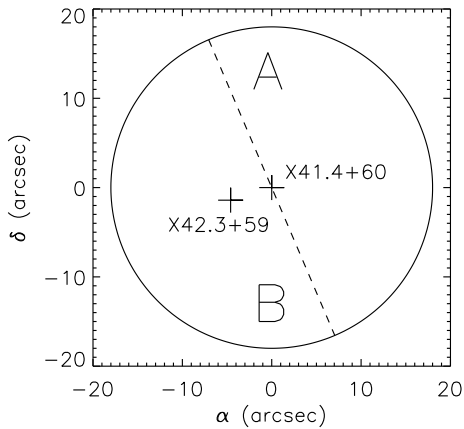}
\caption{ 
The circle indicates a 18\arcsec-radius region around \xa. The dashed line is perpendicular to the direction connecting \xa\ and \xb. Two regions A and B are defined as the two half circles divided by the dashed line, and A+B refers to the whole circular region.
\label{fig:reg}}
\end{figure}

\subsection{Timing Analysis}
\label{sec:timing}
All PN and MOS data were used for the timing analysis in order to increase the signal to noise ratio. We applied FLAG equal to \#XMMEA\_EP and PATTERN between 0 and 4 to select PN events, and FLAG equal to \#XMMEA\_EM and PATTERN between 0 and 12 for MOS events. 

We examined the timing gaps in the data and background flares before binning events into light curves.  CCD timing gaps are stored in the good time interval (GTI) extension of the events file for each CCD chip.  GTIs to exclude background flares are generated on basis of the quiescent intervals in the 10--15 keV light curves. Continuous intervals of source light curves are created from combined events files with a resolution of 0.5~s in the common intervals of all above GTIs.  However, we found the source was so strong that the power spectral density (PSD) shape did not change much whether or not we exclude the background flares. Therefore, we decided not to exclude background flares and only to exclude CCD timing gaps.

Every continuous piece of the light curve longer than 512~s was divided into segments of 512~s each.  Individual PSDs were calculated with a 1024-point fast Fourier transform (FFT) for each 512~s segment.  All PSDs were normalized to rms \citep{van88} and averaged to a final one, which corresponds to an effective exposure of 20.5~ks for the first observation and 50.7~ks for the second. The PSD was rebinned linearly at $\nu\le\nu_{\rm b}$ by a factor of $\delta_1$ and logarithmically at $\nu>\nu_{\rm b}$ by a factor of $\delta_2$. We set $\nu_{\rm b}=0.1$~Hz, $\delta_1=3$ and $\delta_2=1.2$ for the first observation and $\nu_{\rm b}=0.16$~Hz, $\delta_1=5$ and $\delta_2=1.2$ for the second.

\begin{figure}[b]
\centering
\includegraphics[width=0.8\columnwidth]{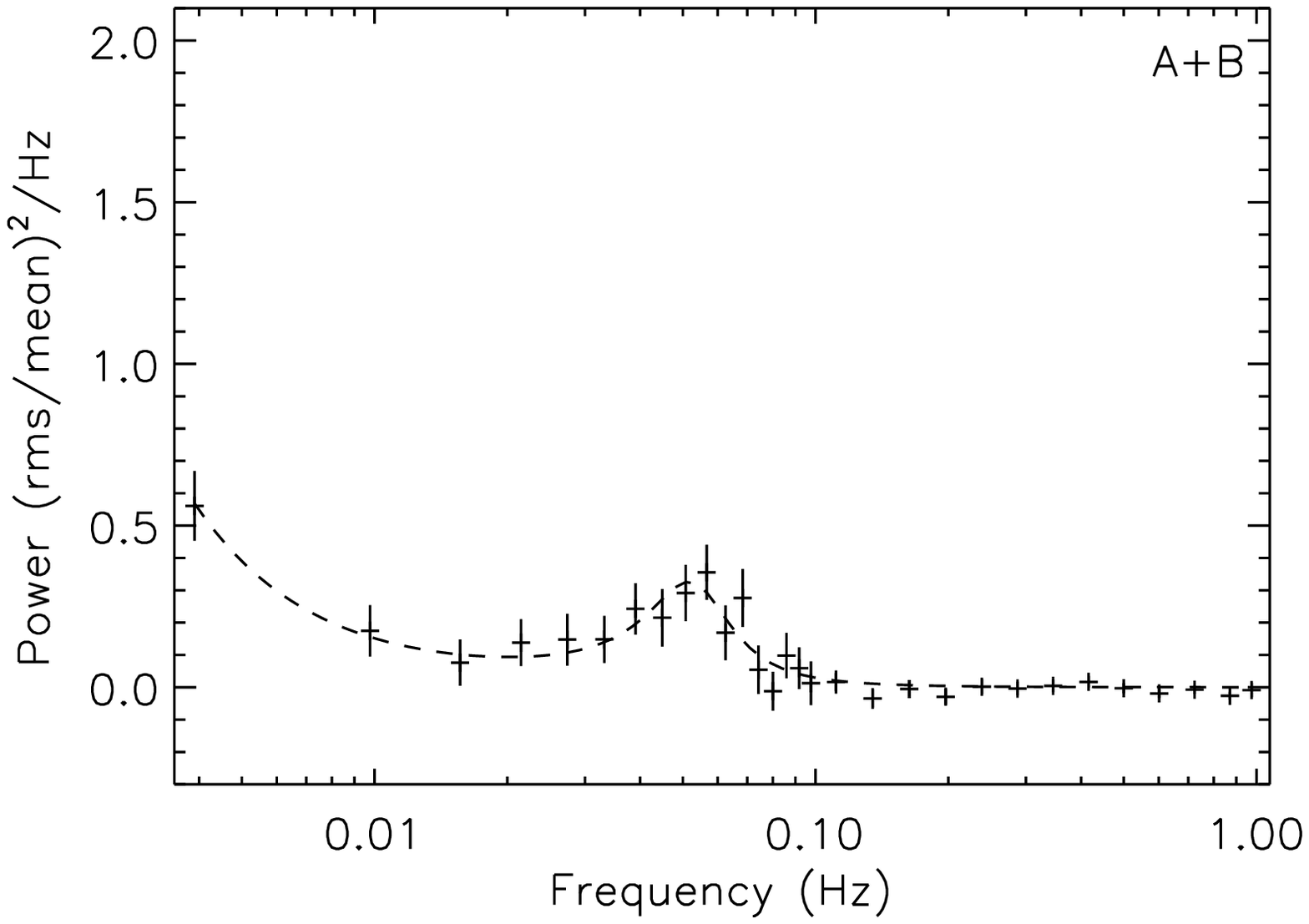}\\
\includegraphics[width=0.8\columnwidth]{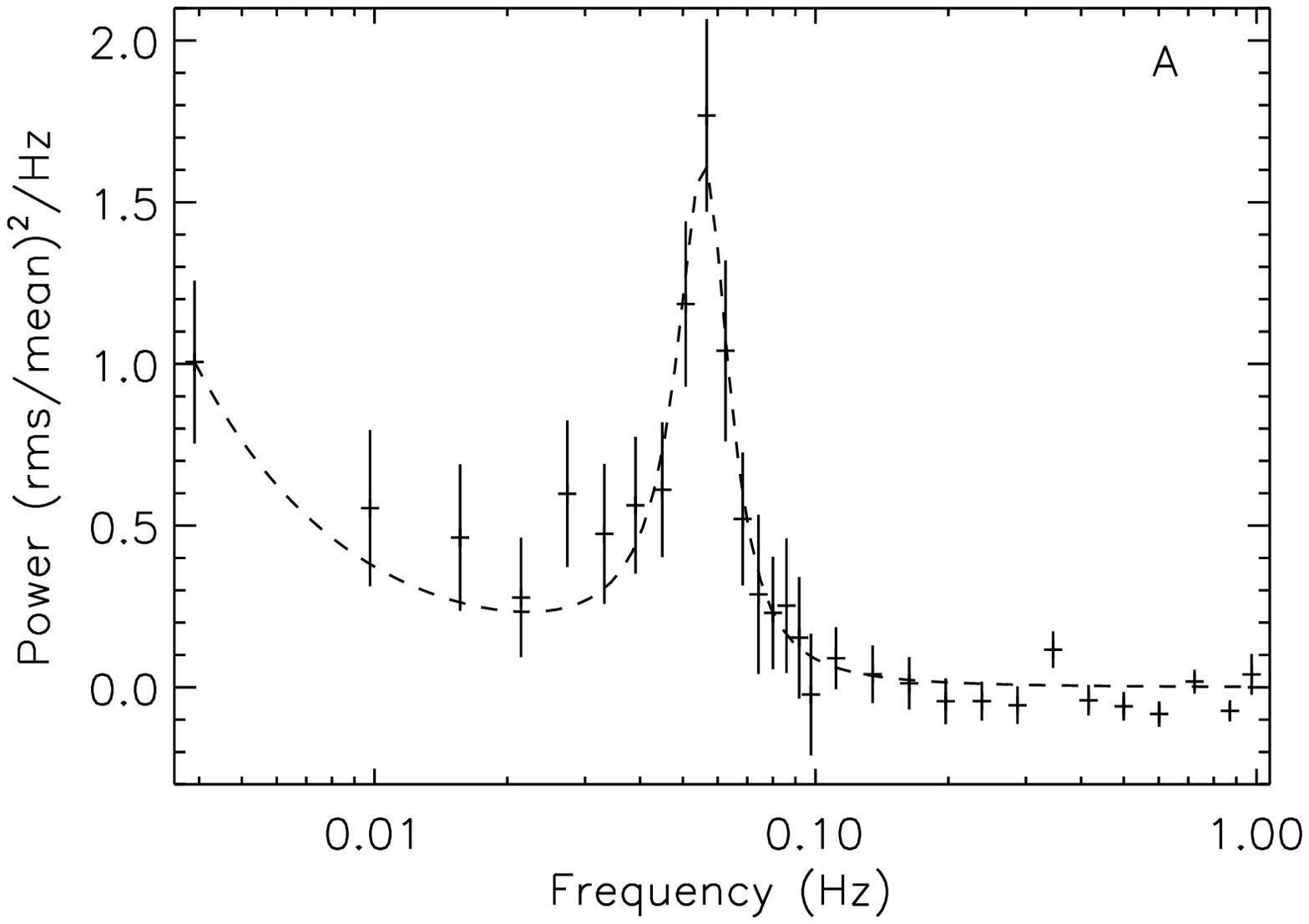}\\
\includegraphics[width=0.8\columnwidth]{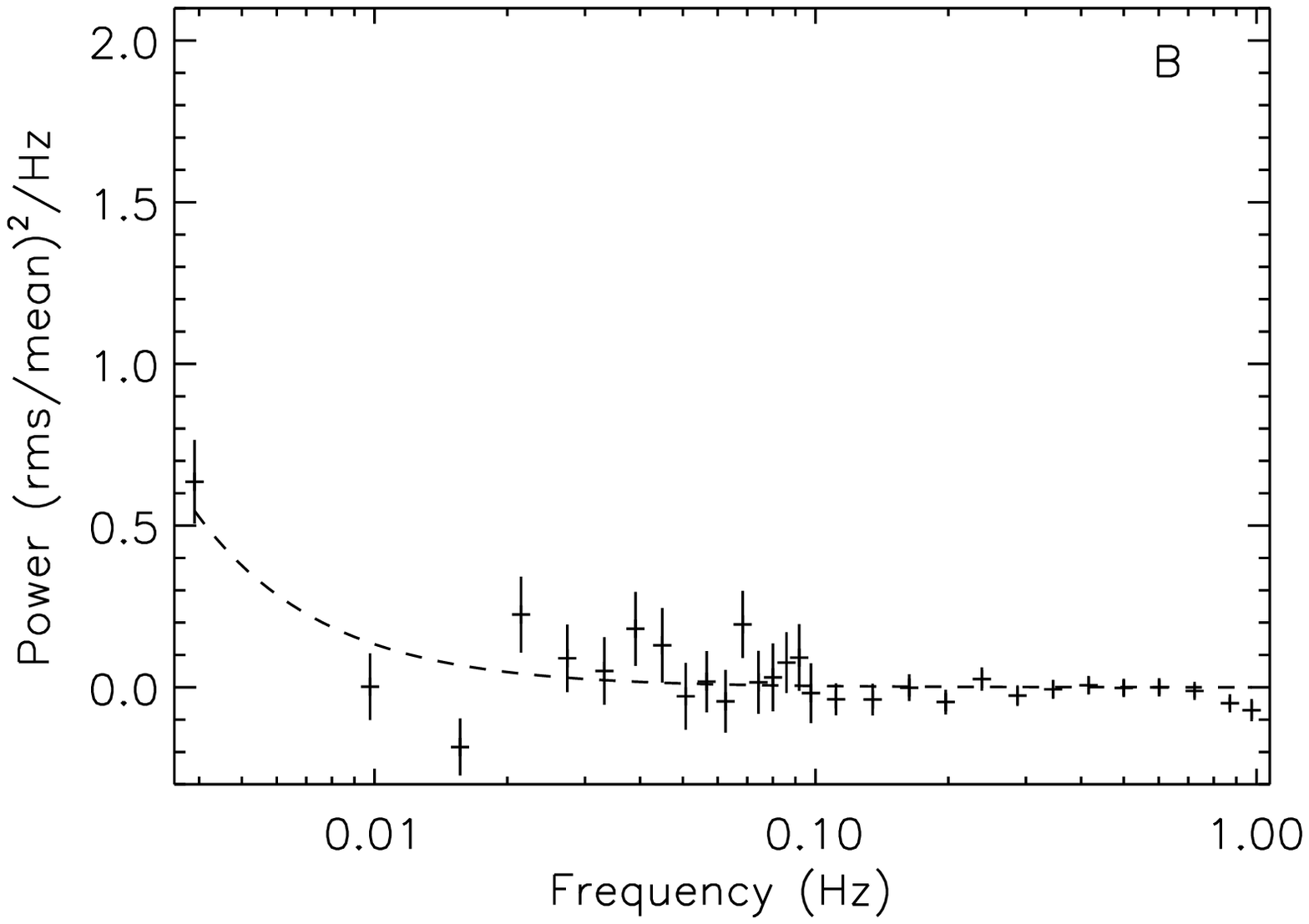}
\caption{
2--10 keV PSD for the first observation calculated from regions A+B, A and B. The dashed line is the best-fit model including a power-law component and a Lorentzian component. The PSD for region B is modeled only by a single power-law.
\label{fig:psd1}}
\end{figure}

\begin{figure}[b]
\centering
\includegraphics[width=0.8\columnwidth]{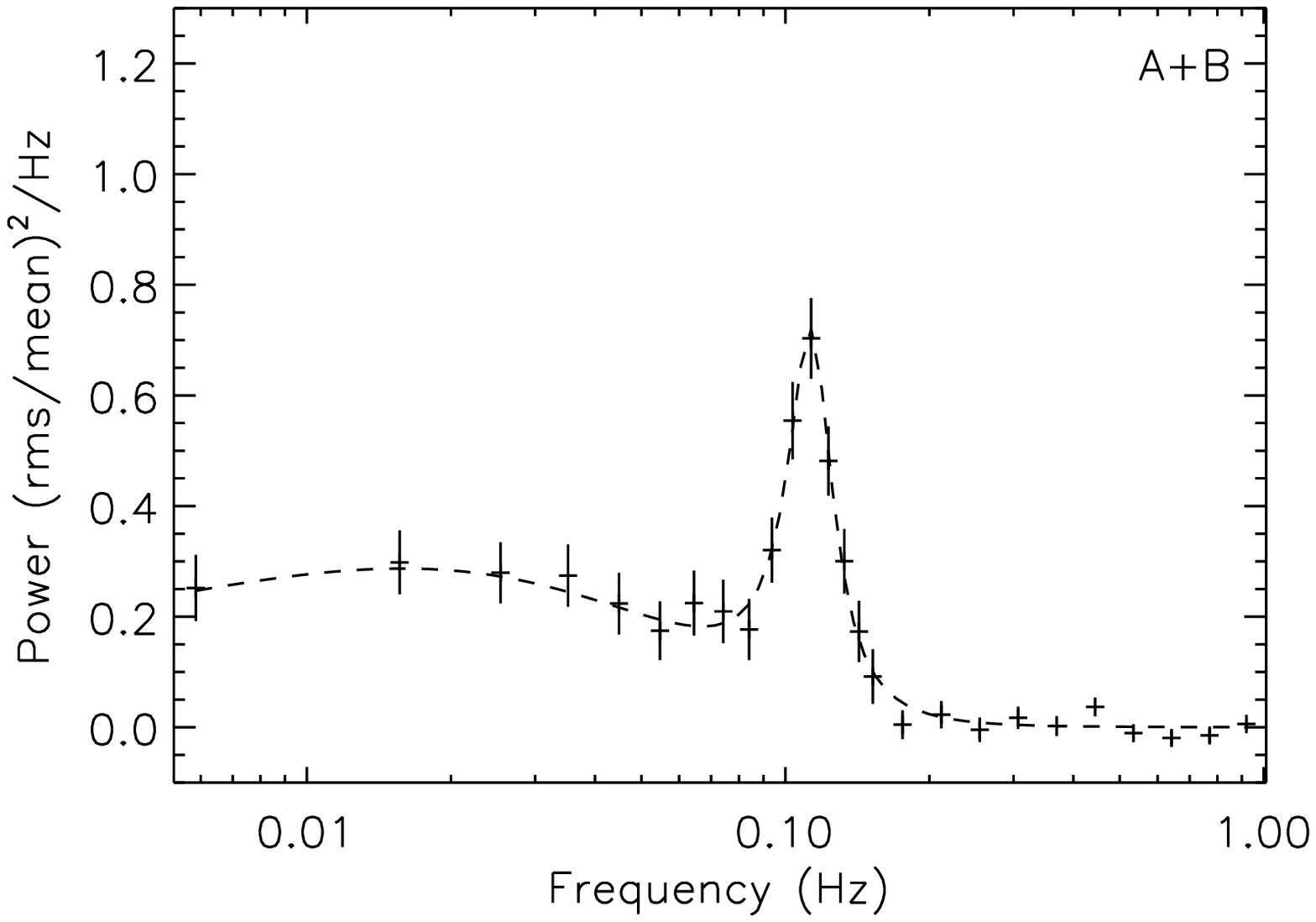}\\
\includegraphics[width=0.8\columnwidth]{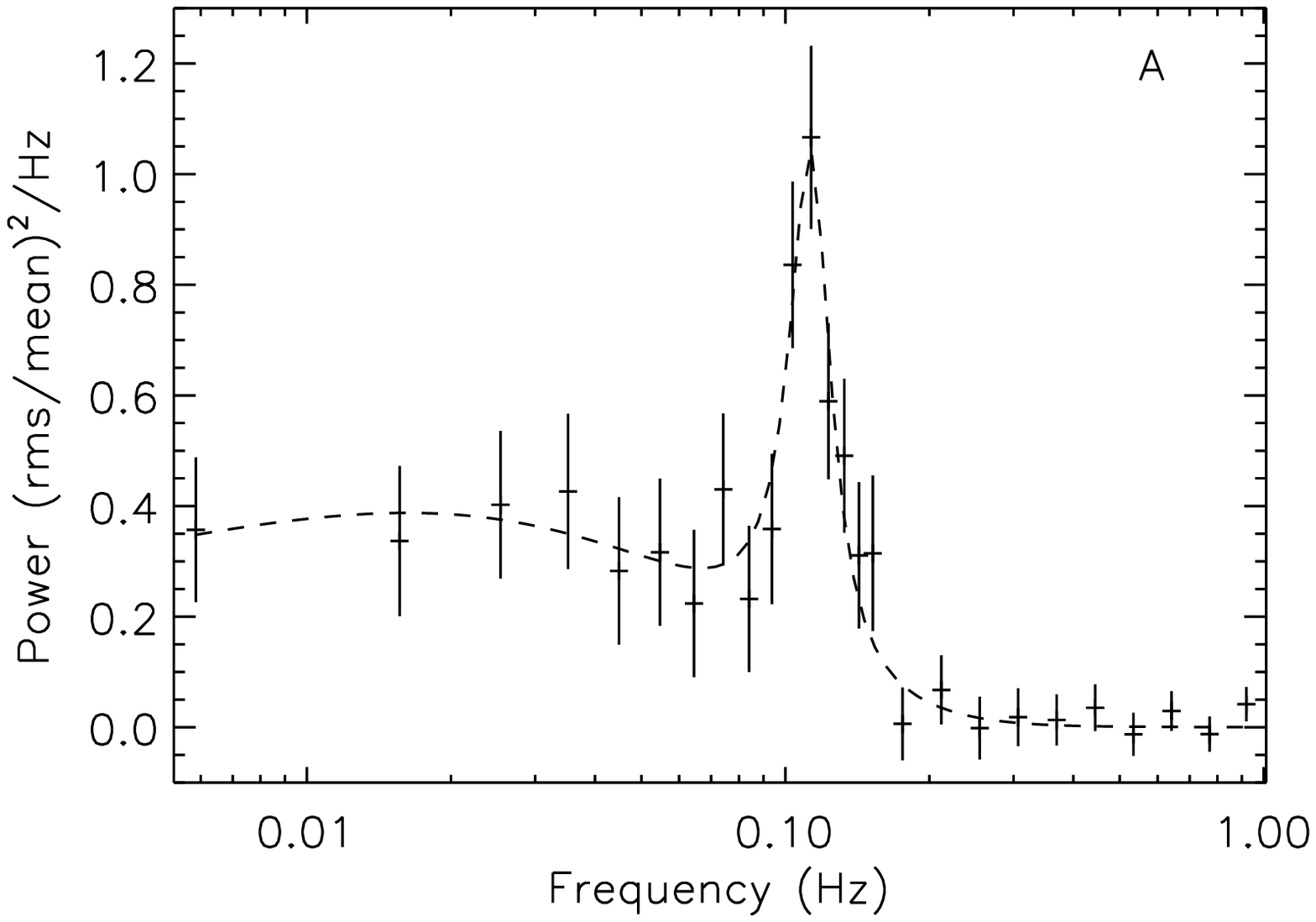}\\
\includegraphics[width=0.8\columnwidth]{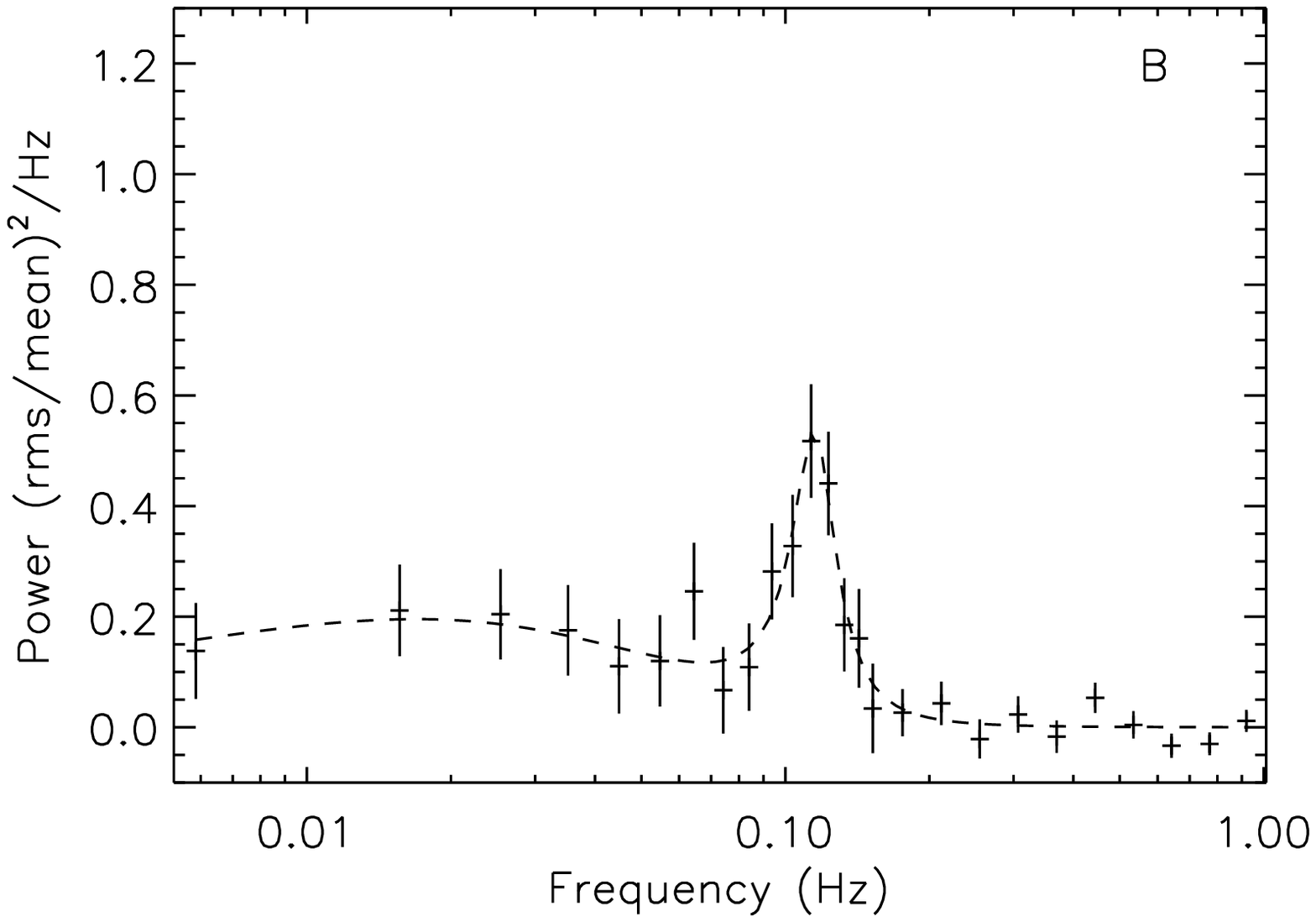}
\caption{
2--10 keV PSD for the second observation calculated from regions A+B, A and B. The dashed line is the best-fit model including a exponentially cutoff power-law component and a Lorentzian component. 
\label{fig:psd2}}
\end{figure}

Three regions defined in Figure~\ref{fig:reg} were used to extract events: an 18\arcsec-radius circle centered on \xa\ divided into two half circles by a line perpendicular to the line connecting \xa\ and \xb. Region A is the half circle on the side without \xb\ and region B is the half circle containing \xb. PSDs for regions A+B, A and B are presented in Figure~\ref{fig:psd1} for the first observation and in Figure~\ref{fig:psd2} for the second.

We fit the PSDs for the first observation with a power-law plus Lorentzian model; the former component is to model the continuum and the latter is for the QPO. For the second observation, the continuum component is modeled by an exponentially cutoff power-law model. The best-fit parameters including the power-law slope $\alpha$, cutoff frequency $\nu_{\rm cut}$, Lorentz centroid $\nu_{\rm L}$ and full width at half maximum (FWHM) $w_{\rm L}$, as well as the fractional rms amplitude (rms/mean) of the QPO component, are listed in Table~\ref{tab:timing} with 1$\sigma$ errors, respectively for each region and observation.  We note that the QPO is not detected in region B for the first observation. We fixed the power-law index to $-$1.5 to model the continuum and evaluated a 2$\sigma$ upper limit of the QPO rms by integrating the powers in the 30--70~mHz frequency region.

Given the resolved source count rates from the surface brightness fits, \xa\ contributes 41\%, 67\% and 30\% of 2--10 keV photons respectively for region A+B, A and B for the first observation, and 84\%, 94\% and 76\% respectively for the second. Assuming the QPO comes from one of the ULXs, the required rms for a particular source is obtained by normalizing the measured rms to the photon fraction in each region. The calculated values are shown in columns (8) and (9) in Table~\ref{tab:timing}. Column (8) shows consistent rms values from the different regions for each observation. Thus, the assumption that \xa\ produces the QPOs gives consistent results regardless of the region chosen to measure the QPO strength. Conversely, column (9) shows that if \xb\ is assumed to produce the QPOs then widely differing rms values are found from the different regions.  Thus, we conclude that the QPOs originate from \xa.

\begin{deluxetable*}{ccccccccc}
\tablewidth{\textwidth}
\tablecolumns{9}
\tablecaption{Best-fit PSD parameters for light curves extracted from different source regions in the 2--10 keV range.
\label{tab:timing}}
\tablehead{
\colhead{region} & \colhead{$\alpha$} & \colhead{$\nu_{\rm cut}$} & \colhead{$\nu_{\rm L}$} & \colhead{$w_{\rm L}$} & \colhead{$\chi^2/$dof} & \colhead{rms$_{\rm QPO}$} & \colhead{rms$_{41.1}$} & \colhead{rms$_{42.2}$} \\
\colhead{} & \colhead{} & \colhead{(mHz)} & \colhead{(mHz)} & \colhead{(mHz)} & \colhead{} & \colhead{} & \colhead{} & \colhead{} \\
\colhead{(1)} & \colhead{(2)} & \colhead{(3)} & \colhead{(4)} & \colhead{(5)} & \colhead{(6)} & \colhead{(7)} & \colhead{(8)} & \colhead{(9)}
}
\startdata
\multicolumn{9}{c}{First observation}\cr
\noalign{\smallskip}\hline\noalign{\smallskip}
A+B & $-1.6\pm0.6$ & \nodata & $52\pm3$     & $29\pm9$ & 21/25 & $12.1\pm0.8$\% & $30\pm2$\% & $20.5\pm1.4$\% \\
A   & $-1.2\pm0.5$ & \nodata & $55.8\pm1.3$ & $19\pm4$ & 26/25 & $21.6\pm0.8$\% & $32.2\pm1.2$\% & $65\pm2$\% \\
B   & $-1.5$ fixed & \nodata & \nodata       & \nodata   & 38/29 & $<11$\%        &$<$37\%&$<$16\%\\
\cutinhead{second observation}
A+B & $0.4\pm0.4$         & $0.03\pm0.02$ & $112.9\pm1.3$ & $30\pm5$ & 14/20 & $17.9\pm0.3$\% & $21.3\pm0.4$\% & $111.9\pm1.9$\%\\
A   & $0.3_{-0.3}^{+0.4}$ & $2_{-2}^{+3}$ & $113\pm2$     & $28\pm8$ & 11/20 & $20.3\pm1.1$\% & $21.6\pm1.2$\% & $338\pm18$\%\\
B   & $0.6_{-0.6}^{+0.9}$ & $3_{-3}^{+14}$& $115\pm3$     & $30\pm9$ & 15/20 & $15.6\pm0.9$\% & $20.5\pm1.2$\% & $65\pm4$\%\\
\enddata

\tablecomments{
Col.~(1): Regions from which the light curves are extracted; see Figure~\ref{fig:reg} for details.
Col.~(2): Exponent of the power-law (first observation) or the exponentially cutoff power-law (second observation).
Col.~(3): Cutoff frequency of the exponentially cutoff power-law (only for second observation).
Col.~(4): Centroid of the Lorentzian.
Col.~(5): Full width at half maximum of the Lorentzian.
Col.~(6): $\chi^2$ and degrees of freedom.
Col.~(7): Fractional rms amplitude of the QPO (= the Lorentzian)
Col.~(8): Required rms if the QPO originates from \xa.
Col.~(9): Required rms if the QPO originates from \xb.
All errors are at 1 $\sigma$ level.}
\end{deluxetable*}

\subsection{A Seven-Year Light Curve of \xb}
\label{sec:x42}

All fourteen available \cha\ observations (see Table~\ref{tab:cha}) and the two \xmm\ observations were used to produce a light curve of the 2--10 keV unabsorbed flux from \xb\ which covers a period of 7.3~years. The \xmm\ observations are described above. The resolved count rates of \xb\ are 0.395~\cts\ on MJD (modified Julian Date) 52035.6 and 0.064~\cts\ on MJD 53117.5.

The source count rate or flux in two \cha\ observations has been
reported in literature.  These are Chandra observation 10, see Table~\ref{tab:cha}, with ACIS \citep{kaa06}, and observation 3 with
HRC \citep{mat01}.

For ACIS observations, we calculated the pileup fraction following the definitions in The \cha\ ABC Guide to Pileup\footnote{http://asc.harvard.edu/ciao/download/doc/pileup\_abc.ps} where $f_t$ is the fraction of good events lost due to grade or energy migration, $f_e$ is the fraction of single events over all detected events, $f_r$ is the fraction of count rate lost, and $\alpha$ is the probability that a piled event is retained as a good grade.

Two ACIS observations, 11 and 12, suffer only mild pileup with $f_t<10\%$. \xb\ has the same count rate in these two observations. Observation 12 is about one day later than observation 11, and has an exposure of only 1/4 of observation 11.  We assume the source stayed in the same state for both observations and use observation 11 to constrain the spectral parameters. The energy spectrum is created from a 1\arcsec\ source region binned for a minimum counts of 100 per bin. We subtracted a background estimated using a nearby diffuse emission region free of point sources and then fitted with a pileup corrected, absorbed power-law model in the energy range of 0.3--8~keV using {\tt Sherpa}. Best-fit parameters include an absorption column density $N_{\rm H}=(3.45\pm0.11)\times 10^{22}$~cm$^{-2}$ and a power-law index $\Gamma=1.49\pm0.06$, with $\chi^2=129$ for 101 degrees of freedom. The pileup model indicates a pileup fraction $f_e=1.2\%$ with $\alpha=0.28$, which is well consistent with the estimate from the count rate. The best-fit model predicts an unabsorbed source flux of $6.0\times10^{-12}$~\ergcms\ in 2--10 keV.  

\begin{deluxetable}{lllll}[t]
\tablecolumns{5}
\tablewidth{\columnwidth}
\tablecaption{\cha\ Observations of M82.
\label{tab:cha}}
\tablehead{
\colhead{index} & \colhead{MJD} & \colhead{ObsID} & \colhead{instrument} & \colhead{exposure}\\
\colhead{} & \colhead{} & \colhead{} & \colhead{} & \colhead{(ks)} \\
\colhead{(1)} & \colhead{(2)} & \colhead{(3)} & \colhead{(4)} & \colhead{(5)}
}
\startdata
1  & 51441.7 & 361 &  ACIS-I & 33.3 \\
2  & 51442.0 & 1302 & ACIS-I & 15.5 \\
3  & 51479.4 & 1411-1 & HRC-I & 36.3 \\
4  & 51542.1 & 378 & ACIS-I & 4.1 \\
5  & 51563.7 & 1411-2 & HRC-I & 17.8 \\
6  & 51614.9 & 379 & ACIS-I & 8.9 \\
7  & 51671.9 & 380-1 & ACIS-I & 3.9 \\
8	& 51707.6 & 380-2 & ACIS-I &  1.2 \\
9  & 52443.9 & 2933 & ACIS-S & 18.0 \\
10 & 53406.3 & 6097 & ACIS-S & 52.8 \\
11 & 53599.5 & 5644 & ACIS-S & 68.1 \\
12 & 53600.8 & 6361 & ACIS-S & 17.5 \\
13 & 54109.7 & 8189 & HRC-S & 61.3 \\
14 & 54112.6 & 8505 & HRC-S & 83.2 \\
\enddata

\tablecomments{ObsID 380 and 1411 consist of two individual observations, respectively. The MJD indicates the mean time of each observation.}
\end{deluxetable}

\begin{figure}[t]
\centering
\includegraphics[width=\columnwidth]{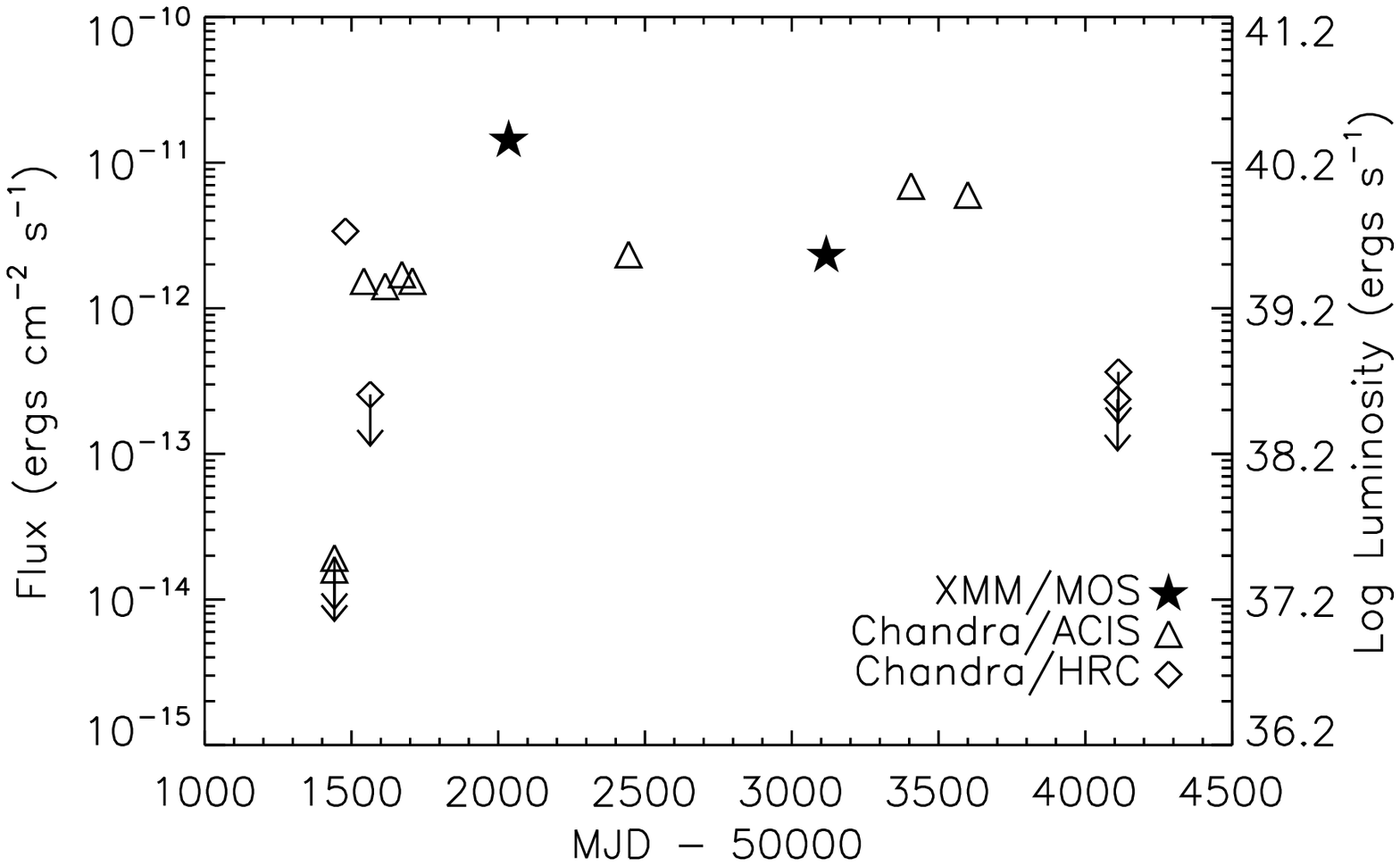}
\caption{
A seven-year light curve of \xb\ observed with \xmm/MOS, \cha/ACIS or \cha/HRC. The left axis indicates the 2--10 keV unabsorbed flux, and the right axis is the corresponding luminosity at a distance of 3.63~Mpc.
\label{fig:x2lc}}
\end{figure}

Five ACIS observations (4, 6, 7, 8, and 9) suffer from severe pileup.  These observations have a pileup fraction $f_t>20\%$ and, thus, are not suitable for spectral analysis.  The count rates are measured from a 2$\times$2 pixel array in 0.3--8 keV and then corrected by $f_r$, which is estimated from a 3$\times$3 pixel island in the full band assuming $\alpha=0.5$. The background level, measured from observation 11, is only about 2\% of the source flux for these observations and thus negligible.

The source was not detected in five observations: three with HRC (5, 13, and 14) and two with ACIS (1 and 2). For the HRC observations, the upper limit of the count rate is calculated from a 1\arcsec\ source region. For the ACIS observations, it is adopted from the brightest pixel around the source region. We note that there is strong diffuse emission around \xb, thus our upper limits are conservative and likely an overestimate. 

All count rates (with a factor of 0.7 for \xmm\ observations) are converted to 2--10 keV unabsorbed flux with PIMMS given $N_{\rm H}=3\times10^{22}$~cm$^{-2}$ and a power-law photon index $\Gamma=1.5$.  The flux estimates from the \xmm\ data include the systematic uncertainty caused by the very nearby sources 4 and 6 as mentioned above. The complete light curve with all the available \xmm\ and \cha\ data is presented in Figure~\ref{fig:x2lc}.  It is clear that \xb\ is highly variable. The source exhibited an unabsorbed flux of $1.4\times10^{-11}$~\ergcms\ and a corresponding luminosity of $2.2\times10^{40}$~\ergs\ in the first \xmm\ observation, reaching its brightest state that has ever been observed.  

\section{Discussion}
\label{sec:dis}

\subsection{Origin of the QPO}

We present solid evidence that the QPOs from M82 originate from the brightest source \xa, rather than the second brightest source \xb.  The angular resolution of \xmm\ is not adequate to cleanly resolve the two sources, but is adequate, with surface brightness fitting using the known source positions, to determine the count rate from each source. Using these count rates and the QPO amplitudes calculated from three different extraction regions, we find that the assumption that \xa\ produces the QPOs give consistent results. Conversely, if \xb\ is taken as the source of the QPOs, then the QPO rms amplitude values calculated from different extraction regions differ at a significance level of at least 20$\sigma$. Therefore, we conclude that the detected QPOs originate from \xa\ at a frequency of 55.8~mHz with an rms of 32\% in the 2--10~keV band in the first observation, and at a frequency of 112.9~mHz with an rms of 21\% in the second observation. 

The origin of the broken power-law feature in the PSD continuum, first reported by \citet{dew06} and \citet{muc06} in the second \xmm\ observation, has not yet been determined.  The same feature is confirmed in our analysis (although we fit this noise component with an exponentially cutoff power-law). The integrated powers for the continuum in the 6--80~mHz range are $14\pm4$\%, $16\pm6$\% and $11\pm5$\%, respectively for region A+B, A and B. Assuming only one of the sources produces the continuum noise, we require an rms amplitude of $17\pm5$\%, $17\pm6$\%, and $14\pm7$\% from \xa, or $87\pm25$\%, $267\pm100$\%, and $46\pm21$\% from \xb, to produce the detected continuum noise in region A+B, A and B, respectively. Because the errors are large, we cannot completely rule out \xb\ as the origin of the continuum noise.  However, it appears that \xa\ is more likely to be the origin of the continuum power.

\subsection{\xa}

It is of interest to use the detected QPO frequencies to attempt to constrain the mass of the compact object in \xa.  Provided the QPO frequency is limited by the Keplerian frequency at the innermost stable circular orbit around a Schwarzschild black hole, the compact object mass of \xa\ is constrained to be less than $3\times10^4 M_\sun$ according to the maximum QPO frequency of 190~mHz that has been reported in \citet{kaa06}. This clearly rules out that \xa\ could be a supermassive black hole.

The centroid of low frequency QPOs in Galactic black holes varies in a large range, e.g., $10^{-2}$--10~Hz for GRS~1915$+$105 \citep{mor97}. Therefore, it is difficult to derive the compact object mass by simply scaling the QPO frequency to mass between ULXs and Galactic black holes.  One must, instead, determine if the QPO frequency correlates with another observable, such as the luminosity or spectral state, which would enable one to correct for the intrinsic frequency variations.

\begin{figure}
\centering
\includegraphics[width=\columnwidth]{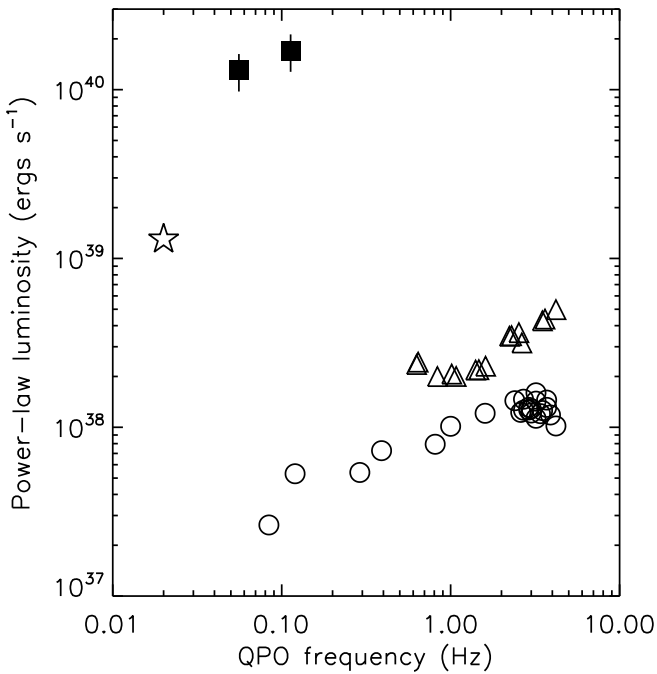}
\caption{
2--10 keV unabsorbed power-law luminosity versus QPO frequency for XTE~J1550$-$564 (circle), GRS~1915$+$105 (triangle), \xa\ (square) and NGC~5408 X-1 (star).
\label{fig:lf}}
\end{figure}

The \xa\ QPO in the second \xmm\ observation is analogous to the ``type
C'' QPO as defined by \citet{rem02}: appearing with a flat-top PSD and
high amplitude.  The QPO in the first \xmm\ observation is also of type
C that has a high amplitude.  A positive correlation between the QPO frequency
and disk/power-law flux has been found for type C QPOs in some stellar mass
black hole binaries like XTE~J1550$-$564 and GRS~1915$+$105 \citep{sob00a,rei00,rem06}. We note that GRO~J1655$-$40 displayed a negative correlation between the QPO frequency and the power-law flux \citep{sob00a}. However, those QPOs have relatively small amplitudes and Q factors are likely not of type C \citep{rem99}. We obtained timing and spectral data for two Galactic sources: XTE~J550$-$564 from
\citet{sob00a,sob00b} and GRS~1915$+$105 from \citet{vig03}. For
XTE~J550$-$564, we consider only QPOs with amplitudes $\geqslant$13\% to
include only type C QPOs. The QPO frequency versus 2-10~keV luminosity
is presented in Figure~\ref{fig:lf}.  For comparison, we plotted the
data for \xa. We include an uncertainty of 25\% on the \xmm\
luminosities for \xa\ to account for uncertainties in the conversion
from XMM-Newton count rates to luminosities due to the lack of knowledge
of the spectrum. The two observations weakly suggest a similar positive
correlation between the QPO frequency and the source flux (power-law
flux).  However, due to the significant uncertainty in the luminosities
extracted from the XMM-Newton data and the fact that there are only two
data points, any conclusions must be regarded as tentative.  We also
plotted one data point for the ULX NGC~5408~X-1, which is the second ULX
found to produce QPOs, in this case near 20~mHz \citep{str07}.  The QPOs
detected from NGC~5408~X-1 are also of type C.

GRS~1915$+$105 is slightly more massive than XTE~J1550$-$564, and its luminosity at a given QPO frequency is slightly higher than XTE~J1550$-$564. If the frequency versus luminosity pattern does, indeed, scale with the mass of the compact object, then the data in Figure~\ref{fig:lf} would indicate that \xa\ and NGC~5408 X-1 are IMBHs.  Though the sample of QPO detections for the ULXs is not large enough to derive the black hole mass accurately, we can see in Figure~\ref{fig:lf} that the pattern for \xa\ is shifted by a factor of 100--500 in luminosity relative to that of the Galactic black holes, indicating a black hole mass of around $10^3M_\sun$. More observations of both ULXs are required to check if the QPO frequency robustly follows a correlation with luminosity similar to that seen from Galactic stellar mass black hole X-ray binaries.

Correlations between the QPO and the power-law photon index have been
observed in several Galactic black holes \citep{sob00a,vig03}.
\citet{tit04} have suggested that the QPO frequency scales inversely
with the mass of the compact object at fixed power-law photon index.
Unfortunately, it is difficult to accurately measure the photon index
for \xa\ or \xb\ with \xmm\ due to large overlap between their PSFs.
This prevents us from making a quantitative comparison with the
relations seen for stellar mass black hole X-ray binaries. It would be
of interest to conduct simultaneous \xmm\ and \cha\ observations in
order to derive timing information from \xmm\ simultaneously with
spectral information from \cha\ in order to study the QPO frequency
versus spectral index correlation for \xa. A combined
spectral/timing analysis of the behavior of \xa\ would likely provide
the most robust estimate of compact object mass.  The same observations
could also greatly improve the uncertainty in luminosity and, thus,
provide a test of the putative QPO frequency versus luminosity
correlation discussed above.

The low frequency break (from slope 0 to $-1$) in the PSD is tightly correlated with the QPO frequency \citep{wij99}. We note the 34~mHz break frequency found in M82~X-1 (plausibly from \xa) and the 113~mHz QPO frequency are consistent with this correlation.  The break frequency in the PSD has been thought to be correlated with the mass of the compact object \citep[e.g.,][]{mar03}. Comparing to Cygnus~X-1, which has a low frequency break varying between 0.02--0.4~Hz \citep{bel90,now99}, the compact object mass of \xa\ would be constrained to be between 6--120 $M_\sun$, assuming a mass of 10$M_\sun$ for Cygnus X-1 \citep{her95}. We note that the estimated mass is lower than that inferred from isotropic emission under the Eddington limit, or the QPO frequency-luminosity correlation discussed above. Again, we emphasize the importance of obtaining multiple observations simultaneously with Chandra and XMM-Newton to study the correlations between spectral and timing properties which could lead to a robust estimate of the compact object mass.

\subsection{\xb}

The seven-year light curve of \xb, see Figure~\ref{fig:x2lc}, shows that the ULX is highly variable.  The peak luminosity implies the compact object mass is at least 200$M_\sun$ assuming isotropic emission.  The ratio of the maximum to minimum observed flux is at least 1000.

The huge variability of \xb\ could be a natural consequence of a relativistic jet aimed along our line of sight -- a so called ``microblazar''.  \xb\ is coincident with the radio source 42.21$+$59.0 \citep{mux94}. However, the radio counterpart presents a thermal spectrum instead of a synchrotron spectrum \citep{mux94,mcd02}, shows little, if any, variability on long time scales \citep{kro00}, and has a spatial extent of 4.9~pc \citep{mcd02}. These results indicate that \xb\ is not relativistic jet emission.

\citet{kal04} suggested an observational test to distinguish whether
ULXs are stellar mass objects with beamed emission or IMBHs. The former
are likely to arise as thermal time-scale mass transfer binaries, in
which the high mass accretion rates lead to thick accretion disks
causing geometrically beamed emission. Thermal time-scale mass transfer
generally produces stable disks and persistent X-ray emission. X-ray
binaries are transient when the disk temperature at outer edge is below
the hydrogen ionization temperature.  This requires an average mass
transfer rate below a critical value.  As first shown by \citet{kin96},
this implies that the black hole mass must be above a minimum set by the
companion star mass and the orbital period.  Transient behavior with a
massive companion star likely requires an IMBH.

Unfortunately, the nature of the \xb\ binary is unknown.  Identification
of the spectral type of the companion star and measurement of the
orbital period, together with the knowledge that the system is an X-ray
transient, could be used to place direct constraints on the mass of the
compact object.  \xb\ is thought to be associated with an H{\sc ii}
region \citep{mux94} that is full of young, hot stars. Simulations
\citep[Fig.~1 and 2 in][]{kal04} show that if the companion was
initially a 10--20$M_\sun$ star, then transient behavior, such as that
observed, likely requires an IMBH.  

From the light curve in Figure~\ref{fig:x2lc}, it seems that \xb\ has just completed an outburst that lasted about 6 years. This outburst duration is longer than that typically seen from Galactic soft X-ray transients, but could be similar to the current outburst of the transient GRS~1915$+$105.  Given an accretion rate ($\dot M$) and donor mass ($M_2$), then according to equation~(2) of \citet{kal04}, an IMBH ($>20M_\sun$) is required for a transient system if the binary period is less than a threshold value
\begin{equation}
P_{\rm trans} \approx
  2.4 \; {\rm yr}
  \left( \frac{\dot M}{10^{-4} \; M_\sun \; {\rm yr}^{-1}} \right)^\frac{1}{1.4}
  \left( \frac{M_2}{10 \; M_\sun} \right)^\frac{1}{7} \; .
\end{equation}
The average accretion rate $\dot M$ can be estimated from the outburst luminosity $L_{\rm out}$ assuming an outburst duty cycle $d$ and an accretion power efficiency $\epsilon$ as $\dot{M}=dL_{\rm out}/(\epsilon c^2)$. We can, thus, rewrite the equation above as
\begin{equation}
P_{\rm trans} \approx
  6.8 \; {\rm d} \times 
  \left( \frac{d \; L_{\rm out}}{\epsilon \; 10^{40} \; {\rm ergs\; s^{-1}} } 
  \right)^\frac{1}{1.4}
  \left( \frac{M_2}{1 \; M_\sun} \right)^\frac{1}{7} \; .
\end{equation}
From Figure~\ref{fig:x2lc}, the outburst luminosity of \xb\ is about $10^{40}$~\ergs. Assuming $d=0.1$ and $\epsilon=0.1$, then $\dot{M} \approx 1.8 \times 10^{-7} M_\sun \; {\rm yr}^{-1}$.  If the companion mass $M_2 \ga 1M_\sun$, then the threshold period for transient behavior for a black hole mass larger than 20$M_\sun$ is 6.8~d. The threshold period depends very weakly on the companion mass. Measurement of the orbital period of \xb\ should enable us to place a lower bound on the compact object mass.

\acknowledgments We thank the anonymous referee for useful comments that
improved our paper. We acknowledge partial support from NASA Grant
NNX06AG77G.  PK acknowledges partial support from a University of Iowa
Faculty Scholar Award.


\begin{thebibliography}{}

\bibitem[Begelman(2002)]{beg02}
Begelman, M. C. 2002, \apj, 568, L97

\bibitem[Belloni \& hasinger(1990)]{bel90}
Belloni, T., \& Hasinger, G. 1990, \aap, 227, L33

\bibitem[Colbert \& Mushotzky(1999)]{col99}
Colbert, E. J. M., \& Mushotzky, R. F. 1999, \apj, 519, 89

\bibitem[Dewangan et al.(2006)]{dew06}
Dewangan, G. C., Titarchuk, L., \& Griffiths, R. E. 2006, \apj, 637, L21

\bibitem[Freedman et al.(1994)]{fre94}
Freedman, W. L. et al. 1994, \apj, 427, 628

\bibitem[Herrero et al.(1995)]{her95}
Herrero, A., Kudritzki, R. P., Gabler, R., Vilchez, J. M., Gabler, A. 1995, \aap, 297, 556

\bibitem[Kaaret et al.(2001)]{kaa01} 
Kaaret, P., Prestwich, A. H., Zezas, A., Murray, S. S., Kim, D.-W., Kilgard, R. E., Schlegel, E. M., Ward, M. J. 2001, \mnras, 321, L29

\bibitem[Kaaret et al.(2006)]{kaa06}
Kaaret, P., Simet, M. G., \& Lang, C. C. 2006, \apj, 646, 174

\bibitem[Kalogera et al.(2004)]{kal04}
Kalogera, V., Henninger, M., Ivanova, N., \& King, A. R. 2004, \apj, 603, L41

\bibitem[King et al.(2001)]{kin01} 
King, A. R., Davies, M. B., Ward, M. J., Fabbiano, G., \& Elvis, M. 2001, \apj, 552, L109

\bibitem[King \& Dehnen(2005)]{kin05}
King, A. R., \& Dehnen, W. 2005, \mnras, 357, 275

\bibitem[King et al.(1996)]{kin96}
King, A.\ R., Kolb, U., \& Burderi, L.\ 1996, \apj,464, L127

\bibitem[K\"ording et al.(2002)]{kor02}  
K\"ording, E., Falcke, H., \& Markoff, S. 2002, \aap, 382, L13

\bibitem[Kronberg et al.(2000)]{kro00}
Kronberg, P. P., Sramek, R. A., Birk, G. T., Dufton, Q. W., Clarke, T. E., \& Allen, M. L. 2000, \apj, 535, 706

\bibitem[Makishima et al.(2000)]{mak00}
Makishima K. et al. 2000, \apj, 535, 623

\bibitem[Markowitz et al.(2003)]{mar03}
Markowitz, A. et al. 2003, \apj, 593, 96

\bibitem[Matsumoto \& Tsuru(1999)]{mat99}
Matsumoto, H.\, \& Tsuru, T.\ 1999, \pasj, 51, 321

\bibitem[Matsumoto et al.(2001)]{mat01}
Matsumoto, H., Tsuru, T. G., Koyama, K., Awaki, H., Canizares, C. R., Kawai, N., Matsushita, S., \& Kawabe, R. 2001, \apj, 547, L25

\bibitem[McDonald et al.(2002)]{mcd02}
McDonald, A.\ R., Muxlow, T.\ W.\ B., Wills, K.\ A., Pedlar, A., Beswick, R.\ J.\ 2002, \mnras, 334, 912 

\bibitem[Morgan et al.(1997)]{mor97}
Morgan, E. H., Remillard, R. A., \& Greiner, J. 1997, \apj, 482, 993

\bibitem[Mucciarelli et al.(2006)]{muc06}
Mucciarelli, P., Casella, P., Belloni, T., Zampieri, L., \& Ranalli, P. 2006, \mnras, 365, 1123

\bibitem[Muxlow et al.(1994)]{mux94}
Muxlow, T. W. B., Pedlar, A., Wilkinson, P. N., Axon, D. J., Sanders, E. M., de Bruyn, A. G. 1994, \mnras, 266, 455

\bibitem[Nowak et al(1999)]{now99}
Nowak, M. A., Vaughan, B. A., Wilms, J., Dove, J. B., Begelman, M. C. 1999, \apj, 510, 847

\bibitem[Poutanen et al.(2007)]{pou07}
Poutanen, J., Lipunova, G., Fabrika, S., Butkevich, A. G., \& Abolmasov, P. 2007, \mnras, 377, 1187

\bibitem[Ptak \& Griffiths(1999)]{pta99}
Ptak, A., \& Griffiths, R. 1999, \apj, 517, L85

\bibitem[Reig et al.(2000)]{rei00}
Reig, P., Belloni, T., van der Klis, M., M\'endez, M., Kylafis, N. D., Ford, E. C. 2000, \apj, 541, 883

\bibitem[Remillard et al.(1999)]{rem99}
Remillard, R.\ A., Morgan, E.\ H., McClintock, J.\ E., Bailyn, C.\ D., Orosz, J.\ A. 1999, \apj, 522, 397

\bibitem[Remillard et al.(2002)]{rem02}
Remillard, R. A., Sobczak, G. J., Muno, M. P., \& McClintock, J. E. 2002, \apj, 564, 962

\bibitem[Remillard \& McClintock(2006)]{rem06}
Remillard, R.\ A., \& McClintock, J.\ E.\ 2006, \araa, 44, 49 

\bibitem[Sobczak et al.(2000a)]{sob00a}
Sobczak, G. J., McClintock, J. E., Remillard, R. A., Cui, W., Levine, A. M., Morgan, E. H., Orosz, J. A., \& Bailyn, C. D. 2000a, \apj, 531, 537

\bibitem[Sobczak et al.(2000b)]{sob00b}
Sobczak, G. J., McClintock, J. E., Remillard, R. A., Cui, W.; Levine, A. M., Morgan, E. H., Orosz, J. A., \& Bailyn, C. D. 2000b, \apj, 544, 993

\bibitem[Strohmayer \& Mushotzky(2003)]{str03}
Strohmayer, T. E., \& Mushotzky, R. F. 2003, \apj, 586, L61

\bibitem[Strohmayer et al.(2007)]{str07}
Strohmayer, T. E., Mushotzky, R. F., Winter, L., Soria, R., Uttley, P., Cropper, M. 2007, \apj, 660, 580

\bibitem[Titarchuk \& Fiorito(2004)]{tit04}
Titarchuk, L., \& Fiorito, R. 2004, \apj, 612, 988

\bibitem[van de Klis(1988)]{van88} 
van der Klis, M. 1988, in Timing Neutron Stars, ed. H. Ogelmen \& E. P. J. van den Heuvel (Dordrecht: Kluwer), 27

\bibitem[Vignarca et al.(2003)]{vig03}
Vignarca, F., Migliari, S., Belloni, T., Psaltis, D., \& van der Klis, M. 2003, \aap, 397, 729

\bibitem[Watarai et al.(2001)]{wat01}
Watarai, K.-Y., Mizuno, T., Mineshige, S. 2001, \apj, 549, L77

\bibitem[Wijnands \& van der Klis(1999)]{wij99}
Wijnands, R., van der Klis, M. 1999, \apj, 514, 939

\end{thebibliography}
\end{document}